\documentclass[11pt,a4paper]{article}
\pdfoutput=1
\usepackage{jheppub}
\usepackage{graphicx}
\usepackage{amssymb}
\usepackage{amsmath}
\usepackage{latexsym}
\usepackage{bbold}
\usepackage{mathbbol}
\newcommand{\de}{\partial} %differential's "de"
\newcommand{\braket}[2]{ \langle #1 \lvert #2 \rangle }  %braket
\newcommand{\ket}[1]{\lvert #1 \rangle} %ket
\newcommand{\bra}[1]{\langle #1 \rvert} %bra
\newcommand{\Braket}[3]{\langle #1 \lvert #3 \rvert #2 \rangle }  %complete braket
\newcommand{\puntos}[1]{^{\bullet} \!\! #1} %left pointed
\newcommand{\puntod}[1]{ #1^{\!\! \bullet} } %right pointed
\newcommand{\puntods}[1]{^{\bullet} \!\! #1^{\!\! \bullet} } %right left pointed
\newcommand{\media}[1]{\langle #1 \rangle} %media
\newcommand{\Det}{\mathrm{Det}}
\newcommand{\Tr}{\mathrm{Tr}}

\newcommand{\M}{\mathcal{M}}
\newcommand{\lie}{\mathit{\pounds}}
\newcommand{\be}{\begin{equation}}
\newcommand{\ee}{\end{equation}}
\newcommand{\bea}{\begin{eqnarray}}
\newcommand{\eea}{\end{eqnarray}}
\title{One-loop quantum gravity from a worldline viewpoint}
\author{Fiorenzo Bastianelli}
\author{and Roberto Bonezzi}
\affiliation{Dipartimento  di Fisica e Astronomia, Universit{\`a} di Bologna and\\
INFN, Sezione di Bologna, via Irnerio 46, I-40126 Bologna, Italy}

\emailAdd{bastianelli@bo.infn.it}\emailAdd{bonezzi@bo.infn.it}

\abstract{

We develop a worldline approach to quantum gravity  in $D=4$.
Using the background field method we consider the covariantly gauge fixed Einstein-Hilbert action
with cosmological constant,  and find a  worldline representation of the
differential operators identified by its quadratic approximation.
We test it by computing the correct one-loop divergencies.
Alternative worldline methods, such as the use of the $O(4)$ spinning particle
that is known to describe correctly the propagation of a massless spin 2 particle in $D=4$,
find obstructions in the coupling to an arbitrary background metric,
apparently preventing a more extensive use in perturbative descriptions of  quantum gravity.
We expect that our model might simplify calculations of  one-loop amplitudes
with respect to standard quantum field theoretical methods.
}


\begin{document}
\maketitle
\flushbottom

\section{Introduction}

Worldline methods have been useful for studying  quantum field theoretical processes both in flat space (see \cite{Schubert:2001he} for a review)
and in curved space (see \cite{Bastianelli:2002fv, Bastianelli:2002qw, Bastianelli:2005vk, Bastianelli:2005uy, Bastianelli:2004zp}
for examples). However, the inclusion of a quantum spin 2 particle to account for the graviton has proved difficult to achieve.
The use of the O(4)-extended spinning particle\footnote{It belongs to the class of O($N$)-extended spinning particles
that describe particles of spin $N/2$ \cite{Gershun:1979fb, Howe:1989vn}.
For $N=0,1,2$ they have been used to describe spin 0, 1/2, and 1 coupled to curved backgrounds in refs.
\cite{Bastianelli:2002fv}, \cite{Bastianelli:2002qw}, and \cite{Bastianelli:2005vk, Bastianelli:2005uy}, respectively.},
which is certainly able to describe the propagation of a massless spin 2 particle in flat space,
has met with severe obstructions in trying to couple it to a generic curved background.
Indeed in \cite{Kuzenko:1995mg}  and \cite{Bastianelli:2008nm} it was found how to couple it to (A)dS and conformally flat spaces,
respectively, but an extension to generic background metrics  remains elusive.

This problem is in a way similar to the case of the O(2)-extended spinning particle, that can be used to describe spin 1 fields coupled
to background gravity, but which does not seem to admit a coupling to background gauge fields.
Nevertheless, it has been used successfully in \cite{Dai:2008bh}
to give a worldline description of non-abelian fields by using BRST and string inspired methods.
The extension of these methods to the  O(4) particle is not obvious.

Given this state of affairs, we try here a simpler, though perhaps less elegant, approach, namely that of deriving from the quadratic
fluctuations  of the gravitational field suitable particle actions that upon quantization give precisely the differential operators of the
quadratic fluctuations. We keep an arbitrary background metric and adopt a convenient background-covariant gauge fixing.
This approach has the virtue of allowing for a useful representation of the one-loop effective action of quantum gravity
in terms of worldline path integrals.
We test this method by reproducing the correct one-loop divergencies for the Einstein gravity \cite{'tHooft:1974bx},
in the presence of the cosmological constant as well \cite{Christensen:1979iy} (see also  \cite{deBerredoPeixoto:2001qx}).

These standard results can be used to test the correctness of the worldline calculation of the heat kernel coefficients in the
one-loop effective action of higher spin fields  on (A)dS spaces, recently performed in \cite{Bastianelli:2012bn} by use of the
O($N$)-extended spinning particles.
One may check that in the case of spin 2 in $D=4$ the coefficient of the logarithmic singularity found in \cite{Bastianelli:2012bn}
differs from the one reported in \cite{Christensen:1979iy}.
However that was not totally unexpected, as one typically finds topological mismatches for dually related theories
in such a coefficient, which is indeed proportional to the Euler character in $D=4$,
see \cite{Duff:1980qv, Bastianelli:2005vk} for the case of differential forms.
The quadratic (and quartic) divergences on the other hand are not topological, but were not reported in
\cite{Christensen:1979iy, deBerredoPeixoto:2001qx}, so that a direct test could not be performed in \cite{Bastianelli:2012bn}.
We find now that they indeed agree with the ones explicitly recomputed with our new method
in  the present paper.

We expect that the flexibility of the worldline path integral representation might be useful in calculating one-loop quantum gravity processes
in a simpler way than standard QFT methods.
Of course, there are many powerful approaches specific to $D=4$ that are being considered in the literature nowadays.
The present work is a step towards the understanding of the role the worldline formalism can play in describing quantized gravitons.
Yet, it would be interesting to find a unified worldline model able to describe the graviton on its own, and not coming directly from manipulating QFT expressions.
The O(4) spinning particle indeed describes spin $2$ massless particles, but it does not allow them to interact with an arbitrary gravitational background, and something new is perhaps needed in that direction.

\section{Background-quantum splitting of Einstein-Hilbert action}

Let us consider a $D$ dimensional curved manifold $\M$ with euclidean signature, equipped with a riemannian metric $G_{\mu\nu}(x)$.
The Einstein-Hilbert action with cosmological constant reads
\begin{equation}\label{EH action}
S[G]=-\frac{1}{\kappa^2}\int d^Dx\,\sqrt{G}\,\Big[R(G)-2\Lambda\Big]\;.
\end{equation}
We perform a background-quantum splitting by defining a fixed classical metric $g_{\mu\nu}(x)$ and quantum metric fluctuations $h_{\mu\nu}(x)$, such that $G_{\mu\nu}(x)=g_{\mu\nu}(x)+h_{\mu\nu}(x)$.
In order to get the effective action at one-loop, it is sufficient to consider the perturbative expansion of \eqref{EH action} up to quadratic order in $h_{\mu\nu}$.
Expanding the Einstein-Hilbert action in powers of $h_{\mu\nu}$ we get
\begin{equation}\label{S(g+h)}
S[g+h]=\frac{1}{\kappa^2}\,\Big[S_0+S_1+S_2+\sum_{n=3}^\infty S_n\Big]\;,
\end{equation}
where
\begin{equation}\label{S012}
\begin{split}
S_0 &= -\int d^Dx\,\sqrt{g}\,\Big\{R-2\Lambda\Big\}\;,\\
S_1 &= \int d^Dx\,\sqrt{g}\,\Big\{h^{\mu\nu}\Big(R_{\mu\nu}-\frac12g_{\mu\nu}R+g_{\mu\nu}\Lambda\Big)\Big\}\;,\\
S_2 &= -\int d^Dx\,\sqrt{g}\,\Big\{\frac14\,h^{\mu\nu}\big(\nabla^2+2\Lambda\big)h_{\mu\nu}-\frac18\,h\big(\nabla^2+2\Lambda\big)h+\frac12\Big(\nabla^\nu h_{\nu\mu}-\frac12\nabla_\mu h\Big)^2\\
&+\frac12\,h^{\mu\lambda}h^{\nu\sigma}\,R_{\mu\nu\lambda\sigma}+\frac12\Big(h^{\mu\lambda}h_\lambda^\nu-\,hh^{\mu\nu}\Big)R_{\mu\nu}+\frac18\Big(h^2-2h^{\mu\nu}
h_{\mu\nu}\Big)R\Big\}\;,
\end{split}
\end{equation}
with $h\equiv g^{\mu\nu}h_{\mu\nu}$.
In the above formulas and from now on all covariant derivatives and curvature tensors are intended to be background quantities built from $g_{\mu\nu}$ only:
$\nabla\equiv\nabla(g)$ and so on. The indices are raised and lowered by means of $g_{\mu\nu}$ and its inverse, so that $h^\mu_\nu\equiv g^{\mu\sigma}h_{\sigma\nu}$ and so forth.

The original action \eqref{EH action} of course enjoys invariance under diffeomorphisms of the full metric
\begin{equation}\label{G diff}
\delta G_{\mu\nu}=\lie_\xi G_{\mu\nu}=\xi^\lambda\de_\lambda G_{\mu\nu}+(\de_\mu\xi^\lambda)G_{\lambda\nu}+(\de_\nu\xi^\lambda)G_{\mu\lambda}
\end{equation}
where $\lie_\xi $ denotes the Lie derivative along the vector field $\xi^ \mu$.
In the split action \eqref{S(g+h)} one can recognize two different symmetries, both taking the same form \eqref{G diff} when acting on the full $G_{\mu\nu}$:
a classical symmetry under background diffeomorphisms of $g_{\mu\nu}$, with $h_{\mu\nu}$ transforming as a tensor, \emph{i.e.}
\begin{equation}\label{BG symmetry}
\delta_\xi\, g_{\mu\nu}=\lie_\xi \,g_{\mu\nu}\;,\quad \delta_\xi \,h_{\mu\nu}=\lie_\xi\, h_{\mu\nu}\;,
\end{equation}
and a quantum gauge symmetry involving $h_{\mu\nu}$ only, that reads
\begin{equation}\label{quantum gauge symmetry}
\begin{split}
\delta_\epsilon\, g_{\mu\nu} &= 0\;,\\
\delta_\epsilon\, h_{\mu\nu} &= \lie_\epsilon\, (g_{\mu\nu}+h_{\mu\nu})=\nabla_\mu\epsilon_\nu+\nabla_\nu\epsilon_\mu+\epsilon^\lambda\nabla_\lambda h_{\mu\nu}+(\nabla_\mu\epsilon^\lambda)h_{\lambda\nu}+(\nabla_\nu\epsilon^\lambda)h_{\mu\lambda}\;.
\end{split}
\end{equation}
%where we recall that $\nabla\equiv\nabla(g)$, and $\epsilon_\mu\equiv g_{\mu\nu}\epsilon^\nu$.

At this point we find convenient to further split the metric fluctuation into traceless and trace parts as
\begin{equation}
h_{\mu\nu}\equiv\bar h_{\mu\nu}+\frac1D g_{\mu\nu}\,h\;,
\end{equation}
with $g^{\mu\nu}\bar h_{\mu\nu}=0$.\footnote{We choose to split the fields into traceless and trace pieces because the operator involved in the effective action for the full fluctuation $h_{\mu\nu}$ would produce, due to the trace projector, a non perturbative vertex in the worldline model that we prefer to avoid.}
The quadratic action in the split form reads
\begin{equation}\label{S quadratic traceless and trace}
S_2 = \int d^Dx\,\sqrt{g}\,\Big\{-\frac14\,\bar h^{\mu\nu}\nabla^2\bar h_{\mu\nu}+\Big(\frac18-\frac{1}{4D}\Big)\,h\nabla^2h-\frac12\Big[\nabla^\nu\bar h_{\nu\mu}+\big(\tfrac{1}{D}-\tfrac12\big)\nabla_\mu h\Big]^2-M_1-M_2\Big\}
\end{equation}
where the curvature pieces $M_1$ and $M_2$ are given by
\begin{equation}\label{M1 M2}
\begin{split}
M_1 &= \frac12\,\bar h^{\mu\lambda}\bar h^{\nu\sigma}\,R_{\mu\nu\lambda\sigma}+\frac12\Big[\bar h^{\mu\lambda}\bar h^\nu_\lambda-\big(1-\tfrac4D\big)h\,\bar h^{\mu\nu}\Big]R_{\mu\nu}-\frac14\,\bar h^{\mu\nu}\bar h_{\mu\nu}\,\big(R-2\Lambda\big)\;,\\
M_2 &= \Big(\frac18-\frac{3}{4D}+\frac{1}{D^2}\Big)\,h^2\,R+\Big(\frac{1}{2D}-\frac14\Big)\,h^2\,\Lambda\;.
\end{split}
\end{equation}
The quantum gauge symmetry for the new fields reads
\begin{equation}\label{gauge symm traceless and trace}
\delta\bar h_{\mu\nu}= \nabla_\mu\epsilon_\nu+\nabla_\nu\epsilon_\mu-\frac2D\,g_{\mu\nu}\nabla\cdot\epsilon+\mathcal{O}(h,\bar h)\;,\quad \delta h=2\,\nabla\cdot\epsilon+\mathcal{O}(h,\bar h)\;.
\end{equation}
We omitted the parts of the gauge transformation \eqref{gauge symm traceless and trace} linear in $h$ and $\bar h_{\mu\nu}$, since they are not needed at one-loop order, but are easily deduced from \eqref{quantum gauge symmetry}.

\section{Gauge fixing}

In order to deal with the action $S_2$ at the quantum level, we have to fix the quantum gauge symmetry \eqref{gauge symm traceless and trace}.
Let us denote
\begin{equation}
f_\mu \equiv \nabla^\nu \bar h_{\nu\mu}+\big(\tfrac1D-\tfrac12\big)\nabla_\mu h\;.
\end{equation}
We choose an $R_\xi$ gauge in order to remove the non laplacian part of the kinetic operator, namely the $f^\mu f_\mu$ part in \eqref{S quadratic traceless and trace}. Hence, using BRST methods, we take as gauge fixing fermion the following functional
\begin{equation}\label{GF fermion}
\Psi=\int d^Dx \sqrt{g}\,b^\mu\,\Big[f_\mu-\frac i2 \pi_\mu\Big]\;,
\end{equation}
where $b^\mu$ is the anti-ghost and $\pi_\mu$ the auxiliary field, such that $\delta_Bb^\mu=i\eta\pi^\mu$, $\delta_B\pi_\mu=0$, $\eta$ being the anticommuting BRST parameter.
The gauge fixing plus ghost actions are obtained now by the BRST variation of \eqref{GF fermion}
with the BRST parameter stripped off
\begin{equation}
S_\pi+S_{\text{gh}} = \frac{\delta_{B}\Psi}{\delta\eta}=\int d^Dx\sqrt{g}\Big\{\Big[\frac{\pi^2}{2}+i\pi^\mu f_\mu\Big]-b^\mu\,s f_\mu\Big\}
\end{equation}
where we denoted $s f_\mu=f_\mu(s\bar h,s h)$. Here $s\bar h_{\mu\nu}$ and $s h$ equal \eqref{gauge symm traceless and trace} with the gauge parameter $\epsilon^\mu$ replaced by the fermionic ghost $c^\mu$ (they correspond to the BRST transformations with the BRST parameter
$\eta$ stripped off).

The path integral over $\pi_\mu$, $\int D\pi\,e^{-S_\pi}$, yields a contribution to the total action that reads $+\frac12\int d^Dx\sqrt{g}\,f^2$, canceling the corresponding term in $S_2$, as can be seen from \eqref{S quadratic traceless and trace}. The total ghost action $S_{\text{gh}}$ consists of a quadratic piece $S_{bc}$ and higher order terms giving $b-c-h$ interactions, that are however irrelevant at the one-loop level. The total action quadratic in the quantum fields thus reads
\begin{equation}\label{Sq}
\begin{split}
S_{q} &= S_{\bar h\,h}+S_{bc}\;,\quad \text{where}\\[2mm]
S_{\bar h\, h} &= \int d^Dx\,\sqrt{g}\,\Big\{-\frac14\,\bar h^{\mu\nu}\nabla^2\bar h_{\mu\nu}+\Big(\frac18-\frac{1}{4D}\Big)\,h\nabla^2h-M_1-M_2\Big\}\;,\\[2mm]
S_{bc}&=-\int d^Dx\,\sqrt{g}\,b^\mu\Big[\nabla^2c_\mu+R_{\mu\nu}\,c^\nu\Big]\;.
\end{split}
\end{equation}
By looking at \eqref{S quadratic traceless and trace} and \eqref{Sq}, one can see that in four dimensions the action for $\bar h_{\mu\nu}$ and $h$ completely separates as $S_{\bar h \,h}=S_{\bar h}+S_{h}$, since the only mixing term $\big(1-\tfrac4D\big)h\,\bar h^{\mu\nu}\,R_{\mu\nu}$ vanishes for $D=4$. After functional integration, this leads to a complete factorization of the corresponding determinants, and for sake of simplicity we will focus from now on only in four dimensions. We set then $D=4$, where $S_{\bar h\, h}$ reads
\begin{equation}\label{Shbarh D=4}
\begin{split}
S_{\bar h\, h} &= \int d^4x\,\sqrt{g}\,\Big\{-\frac14\,\bar h^{\mu\nu}\big(\nabla^2+2\Lambda\big)\bar h_{\mu\nu}+\frac{1}{16}\,h\big(\nabla^2+2\Lambda\big)h\\
&-\frac12\,\bar h^{\mu\lambda}\bar h^{\nu\sigma}\,R_{\mu\nu\lambda\sigma}-\frac12\,\bar h^{\mu\lambda}\bar h^\nu_\lambda\,R_{\mu\nu}-\frac14\,\bar h^{\mu\nu}\bar h_{\mu\nu}\,R\Big\}\;.
\end{split}
\end{equation}

\section{Effective action}

In order to obtain the one-loop effective action, one should integrate $e^{-S_q}$ in $D\bar h\,Dh\,Db\,Dc$. However, at this point one encounters a well-known issue in euclidean quantum gravity: the scalar piece of the action, $S_h$, has the wrong sign in the kinetic term, and this would make the gaussian integral badly divergent. This problem has been widely discussed in the literature, see for instance \cite{Gibbons:1978ac, Mazur:1989by}. We choose to follow the one-loop prescription of Hawking and Gibbons, that suggests to Wick rotate the integration contour in $h$-space to make the integral converge. After doing so, the three actions to be integrated are
\begin{equation}\label{three actions}
\begin{split}
&S_{\bar h} = \int d^4x\,\sqrt{g}\,\Big\{-\frac14\,\bar h^{\mu\nu}\big(\nabla^2+2\Lambda\big)\bar h_{\mu\nu}-\frac12\,\bar h^{\mu\lambda}\bar h^{\nu\sigma}\,R_{\mu\nu\lambda\sigma}-\frac12\,\bar h^{\mu\lambda}\bar h^\nu_\lambda\,R_{\mu\nu}-\frac14\,\bar h^{\mu\nu}\bar h_{\mu\nu}\,R\Big\}\;,\\[2mm]
&S_{h\;\text{rotated}} =-\frac{1}{16}\int d^4x\,\sqrt{g}\,h\big(\nabla^2+2\Lambda\big)h\;,\\[2mm]
&S_{bc}=-\int d^4x\,\sqrt{g}\,b^\mu\Big[\nabla^2c_\mu+R_{\mu\nu}\,c^\nu\Big]\;.
\end{split}
\end{equation}

At this juncture we can write the one-loop partition function for pure gravity by means of the following gauge fixed path integral
\begin{equation}\label{Z[g]}
\begin{split}
Z[g]&=\int D\bar h\,Dh\,Db\,Dc\,e^{-S_q}\\
&\propto\Det^{-1/2}_{TT}\left[K_{\mu\nu,\lambda\sigma}+M_{\mu\nu,\lambda\sigma}\right]\,
\Det^{-1/2}_S\left[-\tfrac12\big(\nabla^2+2\Lambda\big)\right]
\,\Det_V\left[G^\mu_\nu\right]\;,
\end{split}
\end{equation}
where the subscripts of the determinants denote the functional space on which the operators are meant to act, namely symmetric traceless rank two tensors, scalars and vectors.
The operators $K$, $M$ and $G$ are explicitly given by
\begin{equation}\label{K, M and G}
\begin{split}
K_{\mu\nu,\lambda\sigma} &= -\frac14\Big(g_{\mu\lambda}g_{\nu\sigma}+g_{\nu\lambda}g_{\mu\sigma}-\tfrac12\,g_{\mu\nu}g_{\lambda\sigma}\Big)\,\Big(\nabla^2+2\Lambda\Big)\;,\\[2mm]
M_{\mu\nu,\lambda\sigma}&=-\frac12\Big(R_{\mu\lambda\nu\sigma}+R_{\nu\lambda\mu\sigma}-\tfrac12\,g_{\mu\nu}R_{\lambda\sigma}
-\tfrac12g_{\lambda\sigma}R_{\mu\nu}+\tfrac18\,g_{\mu\nu}g_{\lambda\sigma}R\Big)\\
&-\frac14\Big(g_{\nu\sigma}R_{\mu\lambda}+g_{\mu\sigma}R_{\nu\lambda}+g_{\nu\lambda}R_{\mu\sigma}+g_{\mu\lambda}R_{\nu\sigma}-g_{\mu\nu}R_{\lambda\sigma}
-g_{\lambda\sigma}R_{\mu\nu}+\tfrac{1}{4}\,g_{\mu\nu}g_{\lambda\sigma}R\Big)\\
&+\frac14\Big(g_{\mu\lambda}g_{\nu\sigma}+g_{\nu\lambda}g_{\mu\sigma}-\tfrac12\,g_{\mu\nu}g_{\lambda\sigma}\Big)\,R\;,\\[2mm]
G^\mu_\nu &= -\frac12\,\Big(\delta^\mu_\nu\,\nabla^2+R^\mu_\nu\Big)\;.
\end{split}
\end{equation}
The form of the above operators $K$ and $M$ may appear rather cumbersome. This is just because they have to be endomorphisms mapping traceless tensors to traceless tensors. Once they act on the correct functional space, their action greatly simplifies.

The one-loop effective action is defined as $Z[g]=e^{-\Gamma[g]}$ and thus reads
\begin{equation}\label{EA field theory}
\Gamma[g]=\frac12\left\{\Tr_{\scalebox{0.6}{$TT$}}\ln\left[K_{\mu\nu,\lambda\sigma}+M_{\mu\nu,\lambda\sigma}\right]+\Tr_{\scalebox{0.6}{$S$}}\ln\left[-\tfrac12\big(\nabla^2+2\Lambda\big)\right]
-2\Tr_{\scalebox{0.6}{$V$}}\ln\left[G^\mu_\nu\right]\right\}\;.
\end{equation}
In the following section we will study a worldline model able to produce \eqref{EA field theory}.

\section{Worldline representation}

The first step towards the worldline theory is, as usual, the Schwinger proper time representation of the logarithms. Given an operator $\hat{\mathcal{O}}$, one has
\begin{equation}
\Tr\ln\hat{\mathcal{O}}=-\int_0^\infty\frac{d\beta}{\beta}\,\Tr\left[e^{-\beta\hat{\mathcal{O}}}\right]\;.
\end{equation}
The effective action for gravity is thus given by the sum of three traces, that will be reinterpreted as heat kernels of quantum mechanical hamiltonians
\begin{equation}\label{EA proper time}
\Gamma[g]=-\frac12\int_0^\infty\frac{d\beta}{\beta}\left\{\Tr\Big[e^{-\beta(\hat K+\hat M)}\Big]+\Tr\Big[e^{\frac{\beta}{2}(\nabla^2+2\Lambda)}\Big]
-2\Tr\Big[e^{-\beta\hat G}\Big]\right\}\;,
\end{equation}
where the operators $\hat K$, $\hat M$ and $\hat G$ refer to the corresponding ones in \eqref{K, M and G}.
When acting on traceless symmetric tensors and vectors, respectively, they read
\begin{equation}\label{Diff operators}
\begin{split}
(\hat K\,\phi)_{\mu\nu} &= -\frac12\left(\nabla^2+2\Lambda\right)\phi_{\mu\nu}\;,\\[2mm]
(\hat M\,\phi)_{\mu\nu}&=-\frac12\left(R_{\mu\lambda\nu\sigma}+R_{\nu\lambda\mu\sigma}-\tfrac12\,g_{\mu\nu}R_{\lambda\sigma}\right)\,\phi^{\lambda\sigma}
-\frac12\left(R_\mu^\lambda\,\phi_{\lambda\nu}+R_\nu^\lambda\,\phi_{\lambda\mu}-\tfrac12\,g_{\mu\nu}R^{\lambda\sigma}\,\phi_{\lambda\sigma}\right)\\
&+\frac12\,R\,\phi_{\mu\nu}\;,\\[2mm]
(\hat G\,v)_\mu &= -\frac12\Big(\nabla^2v_\mu+R_{\mu\nu}\,v^\nu\Big)\;.
\end{split}
\end{equation}
Since we are going to consider $\hat K+\hat M$, $\hat G$ and $-\tfrac12(\nabla^2+2\Lambda)$ as quantum mechanical hamiltonians, the next task is to construct two
different particle models containing symmetric traceless rank two tensors and vectors in their Hilbert spaces, where the above operators have to be realized. The scalar operator is readily recognized as the
well-known hamiltonian for the bosonic particle in curved space, and needs no further discussion.

\subsection{The traceless tensor model}

Hereby we will study a worldline model able to reproduce $\hat K+\hat M$ as hamiltonian, and containing symmetric traceless rank two tensors in its Hilbert space.
Spacetime coordinates and conjugate momenta $x^\mu(t)$, $p_\mu(t)$ are the phase space bosonic variables of our model. The target space is a curved four dimensional manifold endowed with the metric $g_{\mu\nu}(x)$. To realize our Hilbert space we shall introduce worldline complex fermions that are spacetime symmetric traceless rank two tensors. A few words are now in order. We could indeed use fermionic variables as $\psi^{\mu\nu}(t)$ and $\bar\psi_{\mu\nu}(t)$, but this would lead to several technical issues that are cumbersome to overcome. In fact,
one would  have spacetime dependent $\psi,\bar\psi$ anticommutators.
This choice would produce a lot of ordering issues in fermion bilinears that one would like to avoid. We choose then to introduce a background vielbein $e_\mu^a(x)$ and spin connection $\omega_{\mu ab}(x)$, and we introduce worldline fermions that are spacetime symmetric
 traceless tensors with flat indices: $\psi^{ab}(t)=\psi^{ba}(t)$ and $\bar\psi^{ab}(t)=\bar\psi^{ba}(t)$, obeying $\psi^a_a=\bar\psi^a_a=0$, where $\bar\psi$'s are conjugate momenta of $\psi$'s.
We should stress that, doing this, we are in no way quantizing gravity in the frame formulation. The quantization is performed in metric formulation, as there are no quantum fluctuations of a vielbein, nor a quantum Lorentz gauge symmetry. We are simply introducing a background vielbein, that allows to flatten indices, such as $h_{ab}\equiv e^\mu_a\,e^\nu_b\,h_{\mu\nu}$ and, whenever all operators are covariant objects, one can switch flat to curved indices for free.

Upon canonical quantization these phase space variables obey the following (anti)-commutation relations
\begin{equation}\label{anticommutators}
[x^\mu,p_\nu]=i\,\delta^\mu_\nu\;,\quad \{\psi^{ab},\bar\psi^{cd}\}=\delta^{ac}\delta^{bd}+\delta^{bc}\delta^{ad}-\frac12\,\delta^{ab}\delta^{cd}\;,
\end{equation}
where $\delta^{ab}$ is the flat metric.
We realize them as operators by treating $x^\mu$ and $\psi^{ab}$ as graded coordinates of the wave function, while the momenta are represented as derivatives thereof\footnote{The $g$ factors appearing here ensure hermiticity of $p_\mu$ when using the covariant scalar product, that involves $\sqrt{g}$, and are common when considering quantum mechanics in curved space.
As for the fermions, we find it useful to consider coherent states, as defined in appendix \ref{FCS}, so that any state $\ket{\phi}$
can be represented as a wave function $\phi(x,\psi)= (\bra{x}\otimes \bra{\psi}) \ket{\phi}$.}
\begin{equation}\label{bar psi}
g^{1/4}p_\mu \,g^{-1/4}=-i\de_\mu\;,\quad \bar\psi_{ab}=\frac{\de}{\de\psi^{ab}}\;.
\end{equation}
We stress that, as the trace-free condition is imposed covariantly ( see \eqref{anticommutators}), $\bar\psi^{ab}$ should be represented as a linear combination of $\psi$ derivatives. We represent it formally as \eqref{bar psi}, meaning that $\frac{\de}{\de\psi^{ab}}\,\psi^{cd}=\delta_a^{c}\delta_b^{d}+\delta_b^{c}\delta_a^{d}-\frac12\,\delta_{ab}\delta^{cd}$, that is indeed the identity in trace-free space.
A state in the Hilbert space is represented by a wave function depending on $x$ and $\psi$ variables: $\ket{\phi}\sim\phi(x,\psi)$, where as usual the $\psi$ dependence is polynomial and of finite order. We can then expand the wave function as
\begin{equation}\label{wavefunction}
\ket{\phi}\sim\phi(x,\psi)=\sum_{n=0}^{9}\phi_{(ab)_1...(ab)_n}(x)\psi^{(ab)_1}...\psi^{(ab)_n}\;.
\end{equation}
The sum runs from $0$ to $9$ in four dimensions, but in generic $D$ it would be $\tfrac12(D+2)(D-1)$, the
number of independent components of a rank two symmetric traceless tensor.
We see that, among other fields, we have in the spectrum the rank two symmetric traceless tensor $\phi_{ab}(x)\equiv \bar h_{ab}(x)$ we are looking for. In the following we will project out all the unwanted additional fields so, from now on, we will always consider wave functions containing $\bar h_{ab}$ only: $\ket{\bar h}\sim \bar h_{ab}(x)\psi^{ab}$.

In order to construct covariant derivatives, we introduce Lorentz $SO(D)$ generators
\begin{equation}\label{M generators}
M^{ab}=-M^{ba}:=\frac12[\psi^{ac},\bar\psi_c^b]-\frac12[\psi^{bc},\bar\psi_c^a]=\psi^a\cdot\bar\psi^b-\psi^b\cdot\bar\psi^a\;,
\end{equation}
where we denoted $\psi^a\cdot\bar\psi^b=\psi^{ac}\bar\psi_c^b$. They indeed obey the $SO(D)$ algebra
\begin{equation}\label{so(d) algebra}
[M^{ab},M^{cd}]=\delta^{bc}M^{ad}-\delta^{ac}M^{bd}-\delta^{bd}M^{ac}+\delta^{ad}M^{bc}\;,
\end{equation}
and are used to define the covariant derivative operator as
\begin{equation}\label{covariant der operator}
\hat\nabla_\mu:=\de_\mu+\frac12\,\omega_{\mu ab}\,M^{ab}=\de_\mu+\omega_{\mu ab}\,\psi^a\cdot\bar\psi^b\;.
\end{equation}
We used the hat to distinguish this quantum mechanical operator, acting on wave functions, from the ordinary covariant derivative $\nabla_\mu$ acting on fields. In fact one has
\begin{equation}
\hat\nabla_{\mu}\bar h(x,\psi)= (\nabla_\mu \bar h_{ab})\psi^{ab}=(\de_\mu \bar h_{ab}-\omega_\mu{}^c{}_a\,\bar h_{cb}-\omega_\mu{}^c{}_b\,\bar h_{ac})\psi^{ab}\;.
\end{equation}
Recalling that $p_\mu=-ig^{-1/4}\de_\mu\,g^{1/4}$, the operator can be written in terms of momenta as
\begin{equation}
\hat\nabla_\mu=ig^{1/4}\pi_\mu\,g^{-1/4}=ig^{1/4}\left(p_\mu-i\omega_{\mu ab}\,\psi^a\cdot\bar\psi^b\right)\,g^{-1/4}\;,
\end{equation}
in this way defining the covariant momentum $\pi_\mu$. One should emphasize, however, that further applications of $\hat\nabla$ do not produce covariant objects. For instance, applying it twice one obtains
$$
\hat\nabla_\mu\hat\nabla_\nu \bar h(x,\psi)=(\nabla_\mu\nabla_\nu \bar h_{ab}+\Gamma^\lambda_{\mu\nu}\nabla_\lambda \bar h_{ab})\psi^{ab}\;,
$$
and so on. This is obvious by inspecting \eqref{covariant der operator}, and can be also understood from the fact that $\hat\nabla_\mu\ket{\bar h}$ itself does \emph{not} belong to the Hilbert space. Fortunately, we need only the combination of $\hat\nabla$'s that gives the laplacian, and this is a true endomorphism of the Hilbert space, sending states to states
\begin{equation}\label{Laplacian operator}
\begin{split}
&\hat\nabla^2:=\frac{1}{\sqrt{g}}\hat\nabla_\mu\,g^{\mu\nu}\,\sqrt{g}\,\hat\nabla_\nu=-g^{-1/4}\pi_\mu\,g^{\mu\nu}\,g^{1/2}\pi_\nu\,g^{-1/4}\;,\\[2mm]
&\hat\nabla^2\bar h(x,\psi)=(\nabla^2\bar h_{ab})\psi^{ab}\;.
\end{split}
\end{equation}
With these operators at hand, we can finally give the quantum mechanical representation of the operators $\hat K$ and $\hat M$ of the previous section
\begin{equation}\label{K and M QM}
\begin{split}
\hat K &:= -\frac12\left(\hat\nabla^2+2\Lambda\right)\\
&=\frac12\,g^{-1/4}\pi_\mu\,g^{\mu\nu}\,g^{1/2}\pi_\nu\,g^{-1/4}-\Lambda\;,\\[2mm]
\hat M &:= -\frac12\,R_{abcd}\,\psi^{ac}\bar\psi^{bd}-\frac12\,R_{ab}\,\psi^a\cdot\bar\psi^b+\frac12\,R\;,
\end{split}
\end{equation}
indeed, acting with $\hat M$ on a state one gets
\begin{equation}
\hat M\,\bar h(x,\psi)=-\Big(R_{abcd}\,\bar h^{bd}+R_{ab}\,\bar h^b_c-\tfrac12\,R\,\bar h_{ac}\Big)\psi^{ac}\;,
\end{equation}
that is equivalent to \eqref{Diff operators} if one uses the symmetry and tracelessness of $\psi^{ac}$.
The quantum hamiltonian for our model is thus given by $H=\hat K+\hat M$. To construct a worldline action the only missing piece is a constraint projecting out all the unwanted fields in \eqref{wavefunction} with $n\neq1$. This is easily achieved by constructing the $U(1)$ generator
\begin{equation}\label{U(1) J}
J=\frac14[\psi^{ab},\bar\psi_{ab}]=\mathbf{N}-\frac92\;,
\end{equation}
where $\mathbf{N}$ counts the number of $\psi$'s in a state $\phi(x,\psi)$. Since $J$ commutes with $H$, it can be constrained to have a determined eigenvalue by means of a worldline $U(1)$ gauge field and a Chern-Simons coupling, as it is common in spinning particle models, see \emph{e.g.} \cite{Bastianelli:2005vk}.
The classical action for our model, in phase space and euclidean time is finally given by
\begin{equation}\label{wline action TT phase space}
\begin{split}
S[x,p,\psi,\bar\psi;A] &= \int_0^\beta dt\Big[-ip_\mu\dot x^\mu+\frac12\,\bar\psi_{ab}\dot\psi^{ab}+\frac12\,g^{\mu\nu}\pi_\mu\pi_\nu-\Lambda-\frac12
\,R_{abcd}\,\psi^{ac}\bar\psi^{bd}\\
&-\frac12\,R_{ab}\,\psi^a\cdot\bar\psi^b+\frac12\,R-iA\Big(\frac12\psi^{ab}\bar\psi_{ab}-s\Big)\Big]\;,
\end{split}
\end{equation}
where $A(\tau)$ is a worldine gauge field enforcing $(J-s)\ket{\bar h}=0$. Due to the quantum orderings appearing in \eqref{U(1) J}, the Chern-Simons coupling ensuring $\mathbf{N}=1$ is $s=-\frac72$ or, in arbitrary $D$, $s=1-\frac{(D+2)(D-1)}{4}$. The classical covariant momentum reads $\pi_\mu=p_\mu-i\omega_{\mu ab}\,\psi^a\cdot\bar\psi^b$.
Integrating out momenta $p_\mu$, we get the classical action in configuration space
\begin{equation}\label{wline acrtion TT conf space}
\begin{split}
S[x,\psi,\bar\psi;A] &= \int_0^\beta dt\Big[\,\frac12\,g_{\mu\nu}\dot x^\mu\dot x^\nu+\frac12\,\bar\psi_{ab}\,\big(D_t+iA\big)\,\psi^{ab}
-\frac12\,R_{abcd}\,\psi^{ac}\bar\psi^{bd}\\
&-\frac12\,R_{ab}\,\psi^a\cdot\bar\psi^b+\frac12\,R+iAs-\Lambda\Big]\;,
\end{split}
\end{equation}
where the covariant time derivative for fermions is given by $D_t\psi^{ab}=\dot\psi^{ab}+\dot x^\mu\,(\omega_\mu{}^a{}_c\,\psi^{cb}+\omega_\mu{}^b{}_c\,\psi^{ac})$.

This is a nonlinear sigma model in curved space. It is well known that path integrals for such models need a regularization, due to ill defined products of distributions in the perturbative calculations.
 We shall employ dimensional regularization (DR), widely studied for classes of nonlinear sigma models analogous to the present one \cite{Kleinert:1999aq, Bastianelli:2000pt, Bastianelli:2000nm, Bastianelli:2002qw, Bastianelli:2005vk, Bastianelli:2011cc}. Once the regularization scheme is chosen, one has to identify the correct counterterm to be added to the classical action, to be sure that its path integral corresponds to a precise quantum hamiltonian; in the present situation to $\hat K+\hat M$.
In this case, we may guess that the counterterm has two sources: one is the usual bosonic one, needed to correctly represent $\nabla^2$
(that is $-\frac18 R$ in DR). The other comes from ordering issues of fermions. The path integral usually produces a Weyl ordering (graded-symmetric) for fermionic polynomials. While this does not affect $\omega\psi\bar\psi$ terms (ordering is immaterial due to antisymmetry in $\omega_{\mu ab}$), it produces extra contributions of the form $\xi\,R$ from the potential terms like $\mathbf{Riemann}\,\psi\bar\psi$.
However, to be sure of the correctness of our computations, this educated guess has been checked by directly comparing transition amplitudes obtained with operator methods and path integrals, in close analogy with the procedure employed in \cite{Bastianelli:2011cc}, obtaining
\begin{equation}\label{CT for TT model}
V_{CT}=-\frac78\,R\;,
\end{equation}
where a contribution of $-\frac18\,R$ is the DR bosonic counterterm, while $-\frac34\,R$ is precisely given by Weyl ordering the potential terms in $\hat M$.
It is useful to rescale the time as $t=\beta\tau$, in order to have $\beta$ as a parameter organizing the perturbative expansion. In doing so, we rescale the fermions as $\psi\to\frac{1}{\sqrt\beta}\psi$, $\bar\psi\to\frac{1}{\sqrt\beta}\bar\psi$ and the gauge field as $A\to\frac1\beta A$. We shall also add to \eqref{wline acrtion TT conf space} the counterterm \eqref{CT for TT model}, and we eventually get
\begin{equation}\label{wline action TT rescaled}
\begin{split}
S_{\scalebox{0.5}{TT}}[x,\psi,\bar\psi;A] &=\frac1\beta \int_0^1 d\tau\Big[\,\frac12\,g_{\mu\nu}\dot x^\mu\dot x^\nu+\frac12\,\bar\psi_{ab}\,\big(D_\tau+iA\big)\,\psi^{ab}
-\frac\beta2\,R_{abcd}\,\psi^{ac}\bar\psi^{bd}\\
&-\frac\beta2\,R_{ab}\,\psi^a\cdot\bar\psi^b-\beta^2\Big(\frac38\,R+\Lambda\Big)\Big]+is\int_0^1d\tau\,A\;.
\end{split}
\end{equation}
Before starting the computation of the effective action, we present the worldline model for the ghost sector.

\subsection{The vector model}

In this subsection we will construct a quantum mechanical model having $\hat G$ as hamiltonian operator, and containing vector fields in its Hilbert space.
The graded phase space consists, as usual, of spacetime coordinates and momenta $x^\mu(t)$, $p_\mu(t)$, accompanied by complex worldline fermions $\lambda^a(t)$ and $\bar\lambda^a(t)$, where $a$ is a flat Lorentz index. This construction is similar to the usual $O(2)$ spinning particle \cite{Bastianelli:2005vk, Bastianelli:2005uy}, with the difference that here the states will be vector fields with no gauge symmetry, rather than gauge invariant field strengths.
The canonical (anti-)commutation relations are
\begin{equation}\label{anticommut vector model}
[x^\mu,p_\nu]=i\,\delta^\mu_\nu\;,\quad \{\lambda^{a},\bar\lambda^{b}\}=\delta^{ab}\;.
\end{equation}
Just as before we treat $x^\mu$ and $\lambda^a$ as graded coordinates, and $p_\mu$, $\bar\lambda^a$ as derivatives thereof
$$
g^{1/4}p_\mu \,g^{-1/4}=-i\de_\mu\;,\quad \bar\lambda_{a}=\frac{\de}{\de\lambda^{a}}\;.
$$
The wave function consists now of a set of antisymmetric tensors
\begin{equation}\label{wavefunction vector model}
\ket{\Psi}\sim\Psi(x,\lambda)=\sum_{n=0}^{4}\Psi_{a_1...a_n}(x)\lambda^{a_1}...\lambda^{a_n}\;.
\end{equation}
The vector field we are interested in sits at the $n=1$ level. As we did in the previous case, we are going to project away all the other fields in \eqref{wavefunction vector model} and so we concentrate on states of the form $\ket{v}\sim v_a(x)\,\lambda^a$.
Lorentz generators are readily constructed as $M^{ab}:=\lambda^a\bar\lambda^b-\lambda^b\bar\lambda^a$ and the covariant derivative operator reads
\begin{equation}
\hat\nabla_\mu=ig^{1/4}\pi_\mu\,g^{-1/4}=ig^{1/4}\left(p_\mu-i\omega_{\mu ab}\,\lambda^a\bar\lambda^b\right)\,g^{-1/4}\;.
\end{equation}
The Laplace operator is the same as before when written in terms of covariant momenta
\begin{equation}
\hat\nabla^2:=-g^{-1/4}\pi_\mu\,g^{\mu\nu}\,g^{1/2}\pi_\nu\,g^{-1/4}\;,
\end{equation}
and the full hamiltonian, the operator $\hat G$, is given by
\begin{equation}\label{hamiltonian vector model}
H\equiv \hat G:=\frac12\,g^{-1/4}\pi_\mu\,g^{\mu\nu}\,g^{1/2}\pi_\nu\,g^{-1/4}-\frac12\,R_{ab}\,\lambda^a\bar\lambda^b\;.
\end{equation}
In order to obtain the correct worldline action, we have to project on the $n=1$ sector of the Hilbert space. The $U(1)$ generator counting the tensor rank is
$J:=\frac12[\lambda^a,\bar\lambda_a]=\mathbf{N}-2$ ($\mathbf{N}-\frac D2$ in $D$ dimensions). We gauge the $J$ generator by means of a worldline gauge field $a(t)$, in order to obtain $(J-s)\ket{v}=0$ for a Chern-Simons coupling $s$. Due to the quantum orderings in $J$, the $s$ coupling enforcing $\mathbf{N}=1$ is $s=-1$, or $s=1-\frac D2$ in $D$ dimensions.
With all these ingredients we can write the classical phase space action in euclidean time as
\begin{equation}\label{wline action V phase space}
S[x,p,\lambda,\bar\lambda;a] = \int_0^\beta dt\Big[-ip_\mu\dot x^\mu+\bar\lambda_a\dot\lambda^a+\frac12\,g^{\mu\nu}\pi_\mu\pi_\nu-\frac12\,R_{ab}\,\lambda^a\bar\lambda^b
-ia\Big(\lambda^a\bar\lambda_a-s\Big)\Big]\;,
\end{equation}
where $\pi_\mu=p_\mu-i\omega_{\mu ab}\,\lambda^a\bar\lambda^b$. Integrating out $p_\mu$ we finally get the classical action in configuration space
\begin{equation}\label{wline acrtion V conf space}
S[x,\lambda,\bar\lambda;a] = \int_0^\beta dt\Big[\,\frac12\,g_{\mu\nu}\dot x^\mu\dot x^\nu+\bar\lambda_a\,\big(D_t+ia\big)\,\lambda^a
-\frac12\,R_{ab}\,\lambda^a\bar\lambda^b+ias\Big]\;,
\end{equation}
with the covariant time derivative $D_t\lambda^a=\dot\lambda^a+\dot x^\mu\,\omega_\mu{}^a{}_b\,\lambda^b$.
This nonlinear sigma model has to be regularized as well. We choose again dimensional regularization, and find that the correct counterterm to be added to \eqref{wline acrtion V conf space} is
\begin{equation}\label{CT for V model}
V_{CT}=-\frac38\,R\;.
\end{equation}
As in the tensor model, $-\frac18\,R$ comes from the bosonic part, while $-\frac14\,R$ is due to Weyl ordering of the potential $R_{ab}\lambda^a\bar\lambda^b$.
Also here we find convenient to rescale the time as $t=\beta\tau$, as well as the fermions and the gauge field. Adding also the counterterm \eqref{CT for V model}, we get
\begin{equation}\label{wline action V rescaled}
S_{\scalebox{0.5}{V}}[x,\lambda,\bar\lambda;a] =\frac1\beta \int_0^1 d\tau\Big[\,\frac12\,g_{\mu\nu}\dot x^\mu\dot x^\nu+\bar\lambda_a\,\big(D_\tau+ia\big)\,\lambda^a
-\frac\beta2\,R_{ab}\,\lambda^a\bar\lambda^b-\frac38\,\beta^2R\Big]+is\int_0^1d\tau\, a\;.
\end{equation}

\section{One-loop effective action from worldline path integrals}

We are now ready to put all ingredients together and present the complete worldline representation of
one-loop quantum gravity with cosmological constant. To test it,  we use it to calculate with worldline methods the
first heat kernel coefficients,  also known as Seeley-DeWitt coefficients, that identify the diverging terms of the effective action
(that must be renormalized away to make the effective action finite).

By looking at the representations \eqref{EA field theory} and \eqref{EA proper time} for the effective action, we see that it comes from three separate contributions: from traceless tensor modes, scalar modes, and vector ghosts
\begin{equation}\label{EA splitting}
\Gamma[g]\propto \Gamma_2+\Gamma_0-2\Gamma_1\;,
\end{equation}
where the subscripts refer to rank two traceless tensors, scalars and vectors, respectively.
The Seeley-DeWitt coefficients $a_n(x)$ are identified by expanding the integrands in \eqref{EA proper time} in powers of $\beta$ as
\begin{equation}\label{SDW defined}
\Gamma[g]\propto\int_0^\infty\frac{d\beta}{\beta}\int\frac{d^Dx\sqrt{g(x)}}{(2\pi\beta)^{D/2}}\,\sum_{n=0}^\infty\beta^n\,a_n(x)\;,
\end{equation}
where the prefactor of $\beta^{-D/2}$ is the leading free field behavior.
We will compute separately the three contributions coming from \eqref{EA splitting} by quantizing on the circle the worldline actions presented in the previous section. We perform a perturbative expansion for small $\beta$, that corresponds to the ultraviolet region of the QFT, and compute the Seeley-DeWitt coefficients up to order $\beta^2$, that identify the divergent part of the effective action in four dimensions.
The worldline actions for the tensor and the vector models have a $U(1)$ gauge symmetry that has to be fixed. It is well known, see for instance \cite{Bastianelli:2005vk, Bastianelli:2005uy}, that on the circle the $U(1)$ gauge field can be fixed at most to a constant $\phi$ and one is left, after gauge fixing, with an integral over the modulus $\phi$ corresponding to the gauge invariant Wilson loop: $e^{i\phi}:=e^{i\int_0^1d\tau\,A(\tau)}$. Since the gauge group is abelian, the Faddeev-Popov determinant is just a constant that is absorbed in the overall normalization. After taking this into account we can write down the worldline path integral representations as
\begin{equation}\label{EA Wline}
\begin{split}
\Gamma_2 &= \int_0^\infty\frac{d\beta}{\beta}\int_0^{2\pi}\frac{d\phi}{2\pi}\int_P\mathcal{D}x\int_AD\bar\psi D\psi\, e^{-S_{\scalebox{0.5}{TT}}[x,\bar\psi,\psi;\phi]}\;,\\[2mm]
\Gamma_1 &= \int_0^\infty\frac{d\beta}{\beta}\int_0^{2\pi}\frac{d\theta}{2\pi}\int_P\mathcal{D}x\int_A D\bar\lambda D\lambda \,e^{-S_{\scalebox{0.5}{V}}[x,\bar\lambda,\lambda;\theta]}\;,\\[2mm]
\Gamma_0 &= \int_0^\infty\frac{d\beta}{\beta}\int_P\mathcal{D}x\,e^{-S_{\scalebox{0.5}{S}}[x]}\;,
\end{split}
\end{equation}
where covariant integration over traceless tensor fermions is defined in appendix \ref{FCS}.
Here, $S_{\scalebox{0.5}{TT}}[x,\bar\psi,\psi;\phi]$ and $S_{\scalebox{0.5}{V}}[x,\bar\lambda,\lambda;\theta]$ are the actions \eqref{wline action TT rescaled} and \eqref{wline action V rescaled} with the gauge fields set to constant values: $A(\tau)=\phi$ and $a(\tau)=\theta$. The subscripts $P$ and $A$ stand for periodic and antiperiodic boundary conditions, needed to take traces in the bosonic and fermionic sectors, respectively. $\mathcal{D}x\sim\prod_\tau\sqrt{g\big(x(\tau)\big)}d^4x(\tau)$ denotes the generally covariant measure, while the fermionic measures are flat,
since our fermions are vectors or tensors with flat indices. Finally, the scalar action is simply given by
\begin{equation}\label{scalar action}
S_{\scalebox{0.5}{S}}[x]=\frac1\beta\int_0^1d\tau\Big[\,\frac12\,g_{\mu\nu}\dot x^\mu\dot x^\nu-\beta^2\Big(\frac18\,R+\Lambda\Big)\Big]\;.
\end{equation}
This is the usual bosonic action with DR counterterm, with minimal coupling and a mass term given by $\Lambda$. Its path integral is well known and will not be computed explicitly.

\subsection{Traceless tensor path integral}

We turn now to the perturbative computation of $\Gamma_2$. In order to compute the periodic path integral in $x$, we split the trajectory as $x^\mu(\tau)=x^\mu+q^\mu(\tau)$, where $x^\mu$ is a fixed point in spacetime and the fluctuations vanish at the boundary: $q^\mu(0)=q^\mu(1)=0$. The $x$ path integral then splits as $\int_P\mathcal{D}x=\int d^4x\sqrt{g(x)}\int_D\mathcal{D}q$, where $D$ stands for Dirichlet boundary conditions. To proceed further, following \cite{Bastianelli:1991be, Bastianelli:1992ct},
we get rid of the metric dependent measure $\mathcal{D}q$ by exponentiating the functional product $\prod_\tau\sqrt{g(x(\tau))}$ with bosonic ghosts $a^\mu(\tau)$ and fermionic ghosts $b^\mu(\tau)$, $c^\mu(\tau)$,
such that
$$
\int_D\mathcal{D}q=\int_DDqDaDbDc\,e^{-S_{gh}}\;,\quad\text{where}\quad S_{gh}=\frac1\beta\int_0^1d\tau\,\frac12\,g_{\mu\nu}(x)\Big(a^\mu a^\nu+b^\mu c^\nu\Big)\;.
$$
Having chosen the fixed point $x^\mu$, we expand all the tensor fields around it, in order to perform the perturbative expansion. From $S_{\scalebox{0.5}{TT}}+S_{gh}$ we can now extract the free action
\begin{equation}\label{S2 TT model}
S_2=\frac{1}{2\beta}\,g_{\mu\nu}\int_0^1 d\tau\Big(\dot q^\mu\dot q^\nu+a^\mu a^\nu+b^\mu c^\nu\Big)+\frac{1}{2\beta}\int_0^1d\tau\,\bar\psi_{ab}(\de_\tau+i\phi)\psi^{ab}\;,
\end{equation}
where $g_{\mu\nu}=g_{\mu\nu}(x)$ is the metric at the fixed point.
As usual, from $S_2$ we extract propagators, while the remaining part of the action $S_{int}$ is treated perturbatively in $\beta$. It reads
\begin{equation}\label{Sint TT model}
\begin{split}
S_{int}&=\frac1\beta\int_0^1d\tau\,\Big\{\,\frac12\big[g_{\mu\nu}(x+q)-g_{\mu\nu}(x)\big]\Big(\dot q^\mu\dot q^\nu+a^\mu a^\nu+b^\mu c^\nu\Big)+\omega_{\mu ab}(x+q)\,\dot q^\mu\psi^a\cdot\bar\psi^b\\
&-\frac\beta2\,R_{abcd}(x+q)\,\psi^{ac}\bar\psi^{bd}-\frac\beta2\,R_{ab}(x+q)\,\psi^a\cdot\bar\psi^b-\beta^2\,\frac38\,R(x+q)\Big\}\;,
\end{split}
\end{equation}
where the term with $\Lambda$ and the Chern-Simons part of the action have been omitted, since they completely factorize out. We denote by $\media{...}$ the average computed with the free action $S_2$
\begin{equation}
\media{F}:=\frac{\int_D DqDaDbDc\int_A D\bar\psi D\psi\,F\,e^{-S_2}}{\int_D DqDaDbDc\int_A D\bar\psi D\psi\,e^{-S_2}}\;.
\end{equation}
With this notation, we can write the complete path integral as
\begin{equation}\label{path integral TT model}
\Gamma_2=\int_0^\infty\frac{d\beta}{\beta}\,e^{\beta\Lambda}\int_0^{2\pi}\frac{d\phi}{2\pi}\,\left(2\cos\frac{\phi}{2}\right)^ne^{-is\phi}
\int\frac{d^Dx\sqrt{g(x)}}{(2\pi\beta)^{D/2}}\media{e^{-S_{int}}}\;,
\end{equation}
where $n=\frac{(D+2)(D-1)}{2}$, that is $n=9$ in four dimensions, and $s=1-\frac n2$. The factor $(2\pi\beta)^{-D/2}$ is the usual free bosonic path integral, while $\left(2\cos\frac{\phi}{2}\right)^n$ is the free path integral for $n$ antiperiodic fermions, twisted with the angle $\phi$ as in \eqref{S2 TT model}.

By means of $S_2$ we obtain the following two point functions
\begin{equation}\label{two point functions}
\begin{split}
\media{q^\mu(\tau)q^\nu(\sigma)}&=-\beta\,g^{\mu\nu}\,\Delta(\tau,\sigma)\;,\\
\media{a^\mu(\tau)a^\nu(\sigma)}&=\beta\,g^{\mu\nu}\,\Delta_{gh}(\tau,\sigma)\;,\quad \media{b^\mu(\tau)c^\nu(\sigma)}=-2\beta\,g^{\mu\nu}\,\Delta_{gh}(\tau,\sigma)\;,\\
\media{\psi^{ab}(\tau)\bar\psi^{cd}(\sigma)}&=\beta\,\big(\delta^{ac}\delta^{bd}+\delta^{bc}\delta^{ad}-\tfrac12\,
\delta^{ab}\delta^{cd}\big)\Delta_F(\tau-\sigma,\phi)\;,
\end{split}
\end{equation}
where the inverse metric is intended at the fixed point $x^\mu$, and the (unregulated) propagators are given by
\begin{equation}\label{propagators}
\begin{split}
\Delta(\tau,\sigma)&=\sigma(\tau-1)\theta(\tau-\sigma)+\tau(\sigma-1)\theta(\sigma-\tau)\;,\\
\Delta_{gh}(\tau,\sigma)&=\de_\tau^2\Delta(\tau,\sigma)=\delta(\tau,\sigma)\;,\\
\Delta_F(z,\phi)&=\frac{e^{-i\phi z}}{2\cos\frac{\phi}{2}}\,\Big[e^{i\frac\phi2}\theta(z)-e^{-i\frac\phi2}\theta(-z)\Big]\;,\quad\text{with}\;z=\tau-\sigma\;.
\end{split}
\end{equation}
Here $\delta(\tau,\sigma)$ is the Dirac delta acting on functions vanishing at the boundaries of the segment $[0,1]$. To give a precise meaning to ill defined products and derivatives of these distributions one has to regularize them. We choose, as anticipated, dimensional regularization to compute worldline integrals. A brief discussion of the regularization scheme and more detailed properties of the propagators \eqref{propagators} are contained in appendix \ref{DR app}.
The perturbative calculation can be performed in any coordinate system so, for sake of simplicity, we choose Riemann normal coordinates (and Fock-Schwinger gauge for the spin connection) centered at the fixed point $x^\mu$. Our aim is to compute the Seeley-DeWitt coefficients up to order $\beta^2$, hence the only terms needed in the expansion are the following (obtained, for example, by using the methods of \cite{Muller:1997zk})
\begin{equation}\label{Riemann NC}
\begin{split}
g_{\mu\nu}(x+q)&=g_{\mu\nu}+\frac13\,q^\lambda q^\sigma\,R_{\lambda\mu\nu\sigma}+\mathcal{O}(q^3)+q^\lambda q^\sigma q^\alpha q^\beta\,\Big[\,\frac{1}{20}\nabla_\lambda\nabla_\sigma R_{\alpha\mu\nu\beta}+\frac{2}{45}R_{\tau\lambda\sigma\mu}R^\tau{}_{\alpha\beta\nu}\Big]\\
\omega_{\mu ab}(x+q)&=\frac12\,q^\nu\,R_{\nu\mu ab}+\mathcal{O}(q^2)+q^\nu q^\lambda q^\sigma\,\Big[\,\frac18\,\nabla_\lambda\nabla_\sigma R_{\nu\mu ab}+\frac{1}{24}\,R^\tau{}_{\nu\lambda\mu}R_{\sigma\tau ab}\Big]\\
R_{abcd}(x+q)&= R_{abcd}+\mathcal{O}(q)+\frac12\,q^\mu q^\nu\,\nabla_\mu\nabla_\nu R_{abcd}\;,\\
R_{ab}(x+q)&=R_{ab}+\mathcal{O}(q)+\frac12\,q^\mu q^\nu\,\nabla_\mu\nabla_\nu R_{ab}\;,\quad R(x+q)=R+\mathcal{O}(q)+\frac12\,q^\mu q^\nu\,\nabla_\mu\nabla_\nu R\;,
\end{split}
\end{equation}
where all the tensors on the right hand sides are evaluated at $x$. The terms not written explicitly are those whose path integral is trivially zero, due to the odd number of quantum fields.
We can now use the above expansions in the interacting action \eqref{Sint TT model}. To get terms up to order $\beta^2$ it is sufficient to compute
\begin{equation}\label{pert av}
\media{e^{-S_{int}}}=1-\media{S_4}-\media{S_6}+\frac12\,\media{S_4^2}+\mathcal{O}(\beta^3)\;,
\end{equation}
where $S_n$ refers to the number of quantum fields, and contributes as $\media{S_n}=\mathcal{O}(\beta^{n/2-1})$. Their explicit form is
\begin{equation}\label{S4 S6}
\begin{split}
S_4&= \frac{1}{6\beta}\,R_{\lambda\mu\nu\sigma}\int_0^1d\tau\,q^\lambda q^\sigma\Big(\dot q^\mu\dot q^\nu+a^\mu a^\nu+b^\mu c^\nu\Big)+\frac{1}{2\beta}\,R_{\mu\nu ab}\int_0^1d\tau\,q^\mu\dot q^\nu\psi^a_c\bar\psi^{cb}\\
&-\frac12\,R_{abcd}\int_0^1d\tau\,\psi^{ac}\bar\psi^{bd}-\frac12\,R_{ab}\int_0^1d\tau\,\psi^a_c\bar\psi^{cb}-\frac38\,\beta\,R\;,\\[2mm]
S_6&= \frac1\beta\Big[\,\frac{1}{40}\nabla_\lambda\nabla_\sigma R_{\alpha\mu\nu\beta}+\frac{1}{45}R_{\tau\lambda\sigma\mu}R^\tau{}_{\alpha\beta\nu}\Big]
\int_0^1d\tau\,q^\lambda q^\sigma q^\alpha q^\beta\Big(\dot q^\mu\dot q^\nu+a^\mu a^\nu+b^\mu c^\nu\Big)\\
&+\frac1\beta\Big[\,\frac18\,\nabla_\lambda\nabla_\sigma R_{\mu\nu ab}+\frac{1}{24}\,R^\tau{}_{\mu\lambda\nu}R_{\sigma\tau ab}\Big]\int_0^1d\tau\,q^\lambda q^\sigma q^\mu\dot q^\nu\psi^a_c\bar\psi^{cb}\\
&-\frac14\,\nabla_\mu\nabla_\nu R_{abcd}\int_0^1d\tau\,q^\mu q^\nu\psi^{ac}\bar\psi^{bd}-\frac14\,\nabla_\mu\nabla_\nu R_{ab}\int_0^1d\tau\,q^\mu q^\nu\psi^a_c\bar\psi^{cb}\\
&-\frac{3}{16}\,\beta\,\nabla_\mu\nabla_\nu R\int_0^1d\tau\, q^\mu q^\nu\;.
\end{split}
\end{equation}
We can now compute the quantum averages needed in \eqref{pert av}, by using \eqref{S4 S6} and the propagators given in \eqref{two point functions}, \eqref{propagators}, and we get
\begin{equation}\label{master formula TT}
\begin{split}
\media{e^{-S_{int}}}&=1+\beta\,R\,\left(\frac13+\frac34\,i\tan\frac\phi2\right)\\
&+\beta^2\,\left\{\left(\frac{1}{720}+\frac{1}{16}\cos^{-2}\frac\phi2\right)R^{\mu\nu\lambda\sigma}R_{\mu\nu\lambda\sigma}+\left(-\frac{1}{720}-\frac{3}{16}
\cos^{-2}\frac\phi2\right)R^{\mu\nu}R_{\mu\nu}\right.\\
&\left.+\left(\frac{97}{288}-\frac{7}{32}\cos^{-2}\frac\phi2+\frac{i}{4}\tan\frac\phi2\right)R^2+\left(\frac{7}{240}+\frac{i}{16}\tan\frac\phi2\right)\nabla^2
R\right\}+\mathcal{O}(\beta^3)\;.
\end{split}
\end{equation}
The final step are the modular integrals in \eqref{path integral TT model}, that are easily computed by means of the residue theorem, by setting $z=e^{i\phi}$. Recalling that in four dimensions $n=9$ and $s=-\frac72$, one gets
\begin{equation}\label{modular integrals TT}
\begin{split}
I_1&=\int_0^{2\pi}\frac{d\phi}{2\pi}\left(2\cos\frac\phi2\right)^9e^{i\frac72\phi}=9\;,\\
I_2&=\int_0^{2\pi}\frac{d\phi}{2\pi}\left(2\cos\frac\phi2\right)^9e^{i\frac72\phi}\tan\frac\phi2=7i\;,\\
I_3&=\int_0^{2\pi}\frac{d\phi}{2\pi}\left(2\cos\frac\phi2\right)^9e^{i\frac72\phi}\left(\cos\frac\phi2\right)^{-2}=4\;.
\end{split}
\end{equation}
By using the expansion \eqref{master formula TT} and the integrals above we get the final result for $\Gamma_2$
\begin{equation}\label{SDW TT}
\begin{split}
\Gamma_2=\int_0^\infty\frac{d\beta}{\beta}\int\frac{d^4x\sqrt{g(x)}}{(2\pi\beta)^2}e^{\beta\Lambda}\Big\{9-\frac94\,\beta\,R+\beta^2&\Big[\,\frac{21}{80}
\,R^{\mu\nu\lambda\sigma}R_{\mu\nu\lambda\sigma}-\frac{61}{80}\,R^{\mu\nu}R_{\mu\nu}\\
&+\frac{13}{32}\,R^2-\frac{7}{40}\,\nabla^2R\Big]+\mathcal{O}(\beta^3)\Big\}\;.
\end{split}
\end{equation}
Let us notice that the first Seeley-DeWitt coefficient $a_0$ gives the propagating degrees of freedom. Indeed we obtain $a_0=9$ for a symmetric, rank two, traceless tensor in four dimensions. In order to get the two polarizations of the physical graviton we shall add the contribution of the scalar mode, and subtract the ghosts piece as dictated by \eqref{EA splitting}.
At this point, we can perform the path integral of the vector model in order to obtain $\Gamma_1$.

\subsection{Vector path integral}

The path integral over the circle of the action \eqref{wline action V rescaled} proceeds following the same steps we presented for the tensor model. By splitting the $x$ trajectory, exponentiating the nontrivial measure in $\mathcal{D}q$, and defining averages with the free action, the vector contribution to the effective action can be cast in the form
\begin{equation}\label{path integral V model}
\Gamma_1=\int_0^\infty\frac{d\beta}{\beta}\int_0^{2\pi}\frac{d\theta}{2\pi}\,\left(2\cos\frac{\theta}{2}\right)^De^{-is\theta}
\int\frac{d^Dx\sqrt{g(x)}}{(2\pi\beta)^{D/2}}\media{e^{-S_{int}}}\;,
\end{equation}
with $s=1-\frac D2$. Now fermions are spacetime vectors: $\lambda^a$, $\bar\lambda^a$, hence their contribution in the free path integral is $\left(2\cos\frac\theta2\right)^D$.
The free action used for the averages $\media{...}$ is
\begin{equation}\label{S2 V model}
S_2=\frac{1}{2\beta}\,g_{\mu\nu}\int_0^1 d\tau\Big(\dot q^\mu\dot q^\nu+a^\mu a^\nu+b^\mu c^\nu\Big)+\frac{1}{\beta}\int_0^1d\tau\,\bar\lambda_{a}(\de_\tau+i\theta)\lambda^{a}\;,
\end{equation}
while the interacting action is given by
\begin{equation}\label{Sint V model}
\begin{split}
S_{int}&=\frac1\beta\int_0^1d\tau\,\Big\{\,\frac12\big[g_{\mu\nu}(x+q)-g_{\mu\nu}(x)\big]\Big(\dot q^\mu\dot q^\nu+a^\mu a^\nu+b^\mu c^\nu\Big)+\omega_{\mu ab}(x+q)\,\dot q^\mu\lambda^a\bar\lambda^b\\
&-\frac\beta2\,R_{ab}(x+q)\,\lambda^a\bar\lambda^b-\beta^2\,\frac38\,R(x+q)\Big\}\;.
\end{split}
\end{equation}
We see that the bosonic pieces are identical. All the two point functions not involving fermions $\lambda^a$ and $\bar\lambda^a$ are thus exactly the same of the previous subsection. The two point function for fermions reads
\begin{equation}\label{two point f Vector }
\media{\lambda^a(\tau)\bar\lambda^b(\sigma)}=\delta^{ab}\Delta_F(\tau-\sigma,\theta)\;,
\end{equation}
where $\Delta_F$ is the same as in \eqref{propagators}.
We shall now expand the interaction part \eqref{Sint V model} in Riemann normal coordinates and retain only $S_4$ and $S_6$, that read
\begin{equation}\label{S4 S6 V model}
\begin{split}
S_4&= \frac{1}{6\beta}\,R_{\lambda\mu\nu\sigma}\int_0^1d\tau\,q^\lambda q^\sigma\Big(\dot q^\mu\dot q^\nu+a^\mu a^\nu+b^\mu c^\nu\Big)+\frac{1}{2\beta}\,R_{\mu\nu ab}\int_0^1d\tau\,q^\mu\dot q^\nu\lambda^a\bar\lambda^{b}\\
&-\frac12\,R_{ab}\int_0^1d\tau\,\lambda^a\bar\lambda^{b}-\frac38\,\beta\,R\;,\\[2mm]
S_6&= \frac1\beta\Big[\,\frac{1}{40}\nabla_\lambda\nabla_\sigma R_{\alpha\mu\nu\beta}+\frac{1}{45}R_{\tau\lambda\sigma\mu}R^\tau{}_{\alpha\beta\nu}\Big]
\int_0^1d\tau\,q^\lambda q^\sigma q^\alpha q^\beta\Big(\dot q^\mu\dot q^\nu+a^\mu a^\nu+b^\mu c^\nu\Big)\\
&+\frac1\beta\Big[\,\frac18\,\nabla_\lambda\nabla_\sigma R_{\mu\nu ab}+\frac{1}{24}\,R^\tau{}_{\mu\lambda\nu}R_{\sigma\tau ab}\Big]\int_0^1d\tau\,q^\lambda q^\sigma q^\mu\dot q^\nu\lambda^a\bar\lambda^{b}\\
&-\frac14\,\nabla_\mu\nabla_\nu R_{ab}\int_0^1d\tau\,q^\mu q^\nu\lambda^a\bar\lambda^{b}-\frac{3}{16}\,\beta\,\nabla_\mu\nabla_\nu R\int_0^1d\tau\, q^\mu q^\nu\;.
\end{split}
\end{equation}
Again, the only terms to be computed in the perturbative average are $\media{e^{-S_{int}}}=1-\media{S_4}-\media{S_6}+\frac12\media{S_4^2}$, that give
\begin{equation}\label{master formula V}
\begin{split}
\media{e^{-S_{int}}}&=1+\beta\,R\,\left(\frac13+\frac i4\,\tan\frac\phi2\right)\\
&+\beta^2\,\left\{\left(\frac{1}{720}-\frac{1}{192}\cos^{-2}\frac\phi2\right)R^{\mu\nu\lambda\sigma}R_{\mu\nu\lambda\sigma}+\left(-\frac{1}{720}+\frac{1}{32}
\cos^{-2}\frac\phi2\right)R^{\mu\nu}R_{\mu\nu}\right.\\
&\left.+\left(\frac{25}{288}-\frac{1}{32}\cos^{-2}\frac\phi2+\frac{i}{12}\tan\frac\phi2\right)R^2+\left(\frac{7}{240}+\frac{i}{48}\tan\frac\phi2\right)\nabla^2
R\right\}+\mathcal{O}(\beta^3)\;.
\end{split}
\end{equation}
The modular integrals are easily performed and give (in $D=4$ one has $s=-1$)
\begin{equation}\label{modular integrals V}
\begin{split}
I_1&=\int_0^{2\pi}\frac{d\theta}{2\pi}\left(2\cos\frac\theta2\right)^4e^{i\theta}=4\;,\\
I_2&=\int_0^{2\pi}\frac{d\theta}{2\pi}\left(2\cos\frac\theta2\right)^4e^{i\theta}\tan\frac\theta2=2i\;,\\
I_3&=\int_0^{2\pi}\frac{d\theta}{2\pi}\left(2\cos\frac\theta2\right)^4e^{i\theta}\left(\cos\frac\theta2\right)^{-2}=4\;.
\end{split}
\end{equation}
Putting together \eqref{master formula V} and the modular integrals in \eqref{path integral V model} we finally get the ghost contribution
\begin{equation}\label{SDW V}
\begin{split}
\Gamma_1=\int_0^\infty\frac{d\beta}{\beta}\int\frac{d^4x\sqrt{g(x)}}{(2\pi\beta)^2}\Big\{4+\frac56\,\beta\,R+\beta^2&\Big[\,-\frac{11}{720}
\,R^{\mu\nu\lambda\sigma}R_{\mu\nu\lambda\sigma}+\frac{43}{360}\,R^{\mu\nu}R_{\mu\nu}\\
&+\frac{1}{18}\,R^2+\frac{3}{40}\,\nabla^2R\Big]+\mathcal{O}(\beta^3)\Big\}\;.
\end{split}
\end{equation}

\subsection{Final result}

We are finally ready to put everything together, to obtain the expansion for the gravity effective action. The scalar path integral appearing in \eqref{EA Wline} is well known, as already mentioned, and reads
\begin{equation}\label{SDW S}
\begin{split}
\Gamma_0=\int_0^\infty\frac{d\beta}{\beta}\int\frac{d^4x\sqrt{g(x)}}{(2\pi\beta)^2}e^{\beta\Lambda}\Big\{1+\frac{1}{12}\,\beta\,R+\beta^2&\Big[\,\frac{1}{720}
\,R^{\mu\nu\lambda\sigma}R_{\mu\nu\lambda\sigma}-\frac{1}{720}\,R^{\mu\nu}R_{\mu\nu}\\
&+\frac{1}{288}\,R^2+\frac{1}{120}\,\nabla^2R\Big]+\mathcal{O}(\beta^3)\Big\}\;.
\end{split}
\end{equation}
Assembling the three contributions as dictated by \eqref{EA splitting}: $\Gamma[g]\propto\Gamma_2+\Gamma_0-2\Gamma_1$, we get the final result for the gravity effective action. Expanding in $\beta$ the $\Lambda$ exponentials in $\Gamma_2$ and $\Gamma_0$, it reads
\begin{equation}\label{SDW gravity}
\begin{split}
\Gamma[g]\propto\int_0^\infty\frac{d\beta}{\beta}\int\frac{d^4x\sqrt{g(x)}}{(2\pi\beta)^2}&\Big\{2+\beta\Big(-\frac{23}{6}\,R+10\Lambda\Big)+\beta^2\Big[\,\frac{53}{180}
\,R^{\mu\nu\lambda\sigma}R_{\mu\nu\lambda\sigma}-\frac{361}{360}\,R^{\mu\nu}R_{\mu\nu}\\[2mm]
&+\frac{43}{144}\,R^2-\frac{19}{60}\,\nabla^2R+5\Lambda^2-\frac{13}{6}\,R\Lambda\Big]+\mathcal{O}(\beta^3)\Big\}\;.
\end{split}
\end{equation}
We recognize from $a_0=2$ the physical polarizations of the graviton. By extracting the topological Gauss-Bonnet term $E=R^2_{\mu\nu\lambda\sigma}-4R^2_{\mu\nu}+R^2$, we can rewrite the logarithmic divergent part of $\Gamma[g]$ (the $\mathcal{O}(\beta^2)$
Seeley-DeWitt coefficient) as
\begin{equation}\label{Gamma div gravity}
\begin{split}
\Gamma_{div}[g]=\frac{1}{8\pi^2}\left(\int_{1/M^2}^\infty\frac{d\beta}{\beta}\right)\int d^4x\sqrt{g(x)}\left\{\frac{53}{90}\,E+\frac{7}{20}\,R^{\mu\nu}R_{\mu\nu}+\frac{1}{120}\,R^2+10\Lambda^2-\frac{13}{3}\,R\Lambda\right\}\;,
\end{split}
\end{equation}
where we dropped the total derivative $\nabla^2 R$ and inserted the UV cut-off $\beta\sim {1/M^2}$ in the lower limit of the proper time integral.
 This is the well known result for one-loop divergencies of pure gravity \cite{'tHooft:1974bx, Christensen:1979iy}, and shows that our worldline model correctly reproduce them. 
On AdS the term in curly brackets collapses to the value of $-\frac{571}{2160}R^2$, which may be compared with the results of \cite{Bastianelli:2012bn},
as indicated in the introduction. We conclude that
one may use the present worldline model with confidence  to calculate more demanding one-loop amplitudes in quantum gravity.

%%%%%%%%%%%%%%%%%%%%%%%%%%%%%%%%%%%%%%%%%%%%%%%%%%%%%%%%%%%%%%%%%%
\acknowledgments{The work of FB was supported in part by the MIUR-PRIN contract 2009-KHZKRX.
FB thanks the Galileo Galilei Institute for Theoretical Physics of INFN for hospitality and support extended to him during the completion of this work.
}
%%%%%%%%%%%%%%%%%%%%%%%%%%%%%%%%%%%%%%%%%%%%

\appendix

\section{Fermionic coherent states} \label{FCS}

Here we review some basic formulas about fermionic coherent states, suitably modified for our traceless tensor fermions $\psi^{ab}$ and $\bar\psi^{ab}$.

Given the fermionic operators $\psi^{ab}$ and $\bar\psi^{ab}$, symmetric in $(ab)$ and traceless: $\psi^a_a=\bar\psi^a_a=0$, they obey the oscillator algebra
\begin{equation}
\left\{\psi^{ab},\bar\psi^{cd}\right\}=\delta^{ac}\delta^{bd}+\delta^{bc}\delta^{ad}-\tfrac 2D\,\delta^{ab}\delta^{cd}\;,
\end{equation}
where $a,b=1,...,D$ are flat Lorentz indices. We treat $\psi$'s as creation operators and $\bar\psi$'s as annihilation operators with respect to the vacuum $\ket{0}$. We define coeherent states as
\begin{equation}
\ket{\bar\eta}:=e^{\psi^{ab}\bar\eta_{ab}/2}\ket{0}\;,\quad \bra{\xi}:=\bra{0}\,e^{\xi^{ab}\bar\psi_{ab}/2}\;,
\end{equation}
obeying $\bar\psi^{ab}\ket{\bar\eta}=\bar\eta^{ab}\ket{\bar\eta}$, and the analogous relation for $\bra{\xi}$. They are normalized as
\begin{equation}
\braket{\xi}{\bar\eta}=e^{\xi^{ab}\bar\eta_{ab}/2}\;.
\end{equation}
Since there are $n=\frac{(D+2)(D-1)}{2}$ independent $\psi$'s and $\bar\psi$'s, to define an integral over coherent states
in a Lorentz invariant way,
we introduce Lorentz invariant tensors, built from $\delta_{ab}$ and $\epsilon_{a_1...a_D}$ that we call $Z_{(ab)_1...(ab)_n}$. They have to be symmetric and traceless in each couple $(ab)_i$ and antisymmetric exchanging couples, but we do not need their explicit expressions. For instance, in $D=2$ one has 
$$
Z_{ab\,cd}\propto \epsilon_{ac}\delta_{bd}+\epsilon_{ad}\delta_{bc}+\epsilon_{bd}\delta_{ac}+\epsilon_{bc}\delta_{ad}\;.
$$
The integration measures are then given by
$$
d\xi:=Z_{(ab)_1...(ab)_n}d\xi^{(ab)_1}...d\xi^{(ab)_n}\;,\quad d\bar\eta:=Z_{(ab)_1...(ab)_n}d\bar\eta^{(ab)_n}...d\bar\eta^{(ab)_1}
$$
so that $d\xi d\bar\eta=(-)^n d\bar\eta d\xi$.
The $Z$ tensors are normalized as to give
\begin{equation}\label{cohoerent states formulas}
\begin{split}
&\int d\xi d\bar\eta\,e^{-\tfrac12\,\xi^{ab}\bar\eta_{ab}}=1\\
&\int d\xi d\bar\eta\,e^{-\tfrac12\,\xi^{ab}\bar\eta_{ab}}\ket{\bar\eta}\bra{\xi}=\mathbb{1}\\
& \Tr A=\int d\xi d\bar\eta\,e^{-\tfrac12\,\xi^{ab}\bar\eta_{ab}}\Braket{-\xi}{\bar\eta}{A}=\int d\bar\eta d\xi\,e^{\tfrac12\,\xi^{ab}\bar\eta_{ab}}\Braket{\xi}{\bar\eta}{A} \;.
\end{split}
\end{equation}

\section{Dimensional regularization} \label{DR app}

In this appendix we provide some details on the regularization scheme used in the manuscript, namely dimensional regularization (DR).

The quantum fields of the traceless tensor and vector models have the following two point functions (all the formulas here are valid in four dimensions)
\begin{equation}\label{2 point functions appendix}
\begin{split}
\media{q^\mu(\tau)q^\nu(\sigma)}&=-\beta\,g^{\mu\nu}\,\Delta(\tau,\sigma)\;,\\
\media{a^\mu(\tau)a^\nu(\sigma)}&=\beta\,g^{\mu\nu}\,\Delta_{gh}(\tau,\sigma)\;,\quad \media{b^\mu(\tau)c^\nu(\sigma)}=-2\beta\,g^{\mu\nu}\,\Delta_{gh}(\tau,\sigma)\;,\\
\media{\psi^{ab}(\tau)\bar\psi^{cd}(\sigma)}&=\beta\,\big(\delta^{ac}\delta^{bd}+\delta^{bc}\delta^{ad}-\tfrac12\,
\delta^{ab}\delta^{cd}\big)\Delta_F(\tau-\sigma,\phi)\;,\\
\media{\lambda^a(\tau)\bar\lambda^b(\sigma)}&= \beta\,\delta^{ab}\,\Delta_F(\tau-\sigma,\theta)\;,
\end{split}
\end{equation}
that descend from the quadratic actions
\begin{equation}\label{quad actions appendix}
\begin{split}
S_2^{\scalebox{0.6}{(TT)}}&=\frac{1}{2\beta}\,g_{\mu\nu}\int_0^1 d\tau\Big(\dot q^\mu\dot q^\nu+a^\mu a^\nu+b^\mu c^\nu\Big)+\frac{1}{2\beta}\int_0^1d\tau\,\bar\psi_{ab}(\de_\tau+i\phi)\psi^{ab}\;,\\
S_2^{\scalebox{0.6}{(V)}}&=\frac{1}{2\beta}\,g_{\mu\nu}\int_0^1 d\tau\Big(\dot q^\mu\dot q^\nu+a^\mu a^\nu+b^\mu c^\nu\Big)+\frac{1}{\beta}\int_0^1d\tau\,\bar\lambda_{a}(\de_\tau+i\theta)\lambda^{a}\;.
\end{split}
\end{equation}
The propagators have the continuum limit displayed in \eqref{propagators}, but they come from the mode expansions
\begin{equation}\label{propagators mode expansion}
\begin{split}
&\Delta(\tau,\sigma)=\sum_{m=1}^\infty\left[-\frac{2}{\pi^2\,m^2}\,\sin(\pi m\tau)\sin(\pi m\sigma)\right]=(\tau-1)\sigma\theta(\tau-\sigma)+(\sigma-1)\tau\theta(\sigma-\tau)\;,\\
&\Delta_{gh}(\tau,\sigma)=\sum_{m=1}^\infty 2\,\sin(\pi m\tau)\sin(\pi m\sigma)=\de^2_\tau\Delta(\tau,\sigma)=\delta(\tau,\sigma)\;,\\
&\Delta_F(\tau-\sigma,\phi)=\sum_{r\in\mathbb{Z}+1/2}\frac{-i}{2\pi r+\phi}\,e^{2\pi i r(\tau-\sigma)}=\frac{e^{-i\phi(\tau-\sigma)}}{2\cos\frac{\phi}{2}}\,\Big[e^{i\frac\phi2}\theta(\tau-\sigma)-e^{-i\frac\phi2}\theta(\sigma-\tau)\Big]\;,
\end{split}
\end{equation}
that are easily deduced by expanding the fields with Dirichlet boundary conditions in sine series: $\phi(\tau)=\sum_{m=1}^\infty\phi_m\sin(\pi m\tau)$, and the anti-periodic fermions in half-integer modes: $\Psi(\tau)=\sum_{r\in\mathbb{Z}+1/2}\Psi_r\,e^{2\pi i r\tau}$.
All the above distributions are meant to act on functions defined on the segment $[0,1]$, and vanishing at the boundaries for the case of bosonic
fields and relative ghosts, antiperiodic for the case of fermions.
With the help of \eqref{propagators mode expansion} we can easily find their derivatives and equal time expressions, when well defined:
\begin{equation}\label{propagators derivatives and eq time}
\begin{split}
&\puntos{\Delta}(\tau,\sigma)=\sigma-\theta(\sigma-\tau)\;,\quad\puntod{\Delta}(\tau,\sigma)=\tau-\theta(\tau-\sigma)\;,\\
&\puntods{\Delta}(\tau,\sigma)=1-\delta(\tau,\sigma)\;,\\
&\Delta(\tau,\tau)=\tau(\tau-1)\;,\quad\puntos{\Delta}(\tau,\sigma)\rvert_{\tau=\sigma}=\tau-\frac12\;,
\end{split}
\end{equation}
where dots on the left-right hand side stand for derivatives with respect to the left-right variable.
By symmetric sum of the fermionic series, we also deduce:
\begin{equation}
\Delta_F(0,\phi)=\frac i2\,\tan\frac\phi2\;,
\end{equation}
while, for $\tau\neq\sigma$
\begin{equation}
\Delta_F(\tau-\sigma,\phi)\Delta_F(\sigma-\tau,\phi)=-\frac14\,\cos^{-2}\frac\phi2\;.
\end{equation}

It is well known that products and derivatives of such distributions are ill-defined, and one needs to regularize the path integral. Discretizing the propagation time one gets time slicing regularization, while mode regularization is obtained by cutting off the series in \eqref{propagators mode expansion}. We choose here a different route, dimensional regularization. It consists in continuing the compact time direction with the addition of $n$ non-compact extra dimensions, such that
$$
\tau\in[0,1]\rightarrow t^\alpha=(\tau,\mathbf{t})\in\mathbb{R}^n\times[0,1]\;.
$$
The quadratic action, extended in $n+1$ dimensions reads
\begin{equation}\label{extended DR action}
S_2^{\scalebox{0.6}{(TT)}}=\frac{1}{2\beta}\,g_{\mu\nu}\int_{\mathbb{R}^n\times[0,1]}\!\!\!\!\!\! d^{n+1}t\,\Big( \de_\alpha q^\mu \de^\alpha q^\nu+a^\mu a^\nu+b^\mu c^\nu\Big)+\frac{1}{2\beta}\int_{\mathbb{R}^n\times[0,1]}\!\!\!\!\!\!d^{n+1}t\,\bar\psi_{ab}(\gamma^\alpha\de_\alpha+i\phi)\psi^{ab}\;,
\end{equation}
and similarly for the vector model, where $\de_\alpha=\frac{\de}{\de t^\alpha}$, and $\gamma^\alpha$ are the gamma matrices in $n+1$ dimensions.
The propagators in extended space are given by
\begin{equation}\label{propagators extended}
\begin{split}
\Delta(t,s)&=\int\frac{d^nk}{(2\pi)^n}\sum_{m=1}^\infty\frac{-2}{(\pi m)^2+\mathbf{k}^2}\,\sin(\pi m\tau)\sin(\pi m\sigma)\,e^{i\mathbf{k}\cdot(\mathbf{t}-\mathbf{s})}\;,\\[3mm]
\Delta_{gh}(t,s)&=\int\frac{d^nk}{(2\pi)^n}\sum_{m=1}^\infty2\sin(\pi m\tau)\sin(\pi m\sigma)\,e^{i\mathbf{k}\cdot(\mathbf{t}-\mathbf{s})}\\
&=\delta(\tau,\sigma)\delta^n(\mathbf{t}-\mathbf{s})\;,\\[3mm]
\Delta_F(t-s,\phi)&=-i\int\frac{d^nk}{(2\pi)^n}\sum_{r\in\mathbb{Z}+1/2}\frac{2\pi r\gamma^0+\mathbf{k}\cdot\mathbf{\gamma}-\phi}{(2\pi r)^2+\mathbf{k}^2-\phi^2}\,e^{2\pi ir(\tau-\sigma)}e^{i\mathbf{k}\cdot(\mathbf{t}-\mathbf{s})}\;.
\end{split}
\end{equation}
The propagators in extended space obey the following Green's equations
\begin{equation}\label{Green eqs}
\begin{split}
\de^\alpha\de_\alpha\Delta(t,s)=\Delta_{gh}(t,s)=\delta(\tau,\sigma)\delta^n(\mathbf{t}-\mathbf{s})\;,\\
\left(\gamma^\alpha\frac{\de}{\de t^\alpha}+i\phi\right)\Delta_F(t-s,\phi)=\delta_A(\tau-\sigma)\delta^n(\mathbf{t}-\mathbf{s})\;,
\end{split}
\end{equation}
where $\delta_A$ is the delta distribution acting on antiperiodic functions on $[0,1]$.
Another useful identity, that can be obtained in $n+1$ dimensions, reads
\begin{equation}\label{identity}
\left.\left[\left(\frac{\de^2}{\de t^\alpha\de s_\alpha}+\frac{\de^2}{\de t^\alpha\de t_\alpha}\right)\Delta(t,s)\right]\right\rvert_{t=s}=\frac{\de}{\de\tau}\left.\left[\left(\frac{\de}{\de\tau}\Delta(t,s)\right)\right\rvert_{t=s}\right]\;.
\end{equation}
In practical calculations, one does not employ the extended propagators \eqref{propagators extended}. If the worldline diagram is ill-defined, one extends it with the extra dimensions of DR. In the extended space it is safe to integrate against the delta distributions and to integrate by parts, taking advantage of the non-compact dimensions and, using $\eqref{Green eqs}$, \eqref{identity} and integrations by parts, one can recast the integrals in a form that is not ambiguous anymore, namely with no products of delta distributions nor divergent quantities such as $\delta(\tau,\tau)$. At this point it is possible to remove the regularization, and the result is computed in the original one dimensional form. The ghost system $(a,b,c)$ ensures that every apparently divergent piece (coming from $\dot q\dot q$ propagators) is removed after doing the allowed manipulations.

Since the fermion propagator, twisted with the $U(1)$ modulus $\phi$, is exactly the same as in \cite{Bastianelli:2005vk}, there are no new diagrams to be regulated, and we refer to \cite{Bastianelli:2000nm, Bastianelli:2000dw, Bastianelli:2002qw, Bastianelli:2005vk, Bastianelli:2006rx} for explicit examples of actual computations in DR.

%%%%%%%%%%%%%%%%%%%%%%%%%%%%%%%%%%%%%%%%%%%%%%%%%%%%%


\begin{thebibliography}{99}

%\cite{Schubert:2001he}
\bibitem{Schubert:2001he}
  C.~Schubert,
  ``Perturbative quantum field theory in the string inspired formalism,''
  Phys.\ Rept.\  {\bf 355} (2001) 73
  [hep-th/0101036].
  %%CITATION = HEP-TH/0101036;%%

%\cite{Bastianelli:2002fv}
\bibitem{Bastianelli:2002fv}
  F.~Bastianelli and A.~Zirotti,
  ``Worldline formalism in a gravitational background,''
  Nucl.\ Phys.\ B {\bf 642} (2002) 372
  [hep-th/0205182].
  %%CITATION = HEP-TH/0205182;%%

%\cite{Bastianelli:2002qw}
\bibitem{Bastianelli:2002qw}
  F.~Bastianelli, O.~Corradini and A.~Zirotti,
  ``Dimensional regularization for N=1 supersymmetric sigma models and the worldline formalism,''
  Phys.\ Rev.\ D {\bf 67}, 104009 (2003)
  [hep-th/0211134].
  %%CITATION = HEP-TH/0211134;%%

%\cite{Bastianelli:2005vk}
\bibitem{Bastianelli:2005vk}
  F.~Bastianelli, P.~Benincasa and S.~Giombi,
  ``Worldline approach to vector and antisymmetric tensor fields,''
  JHEP {\bf 0504} (2005) 010
  [hep-th/0503155].
  %%CITATION = HEP-TH/0503155;%%

%\cite{Bastianelli:2005uy}
\bibitem{Bastianelli:2005uy}
  F.~Bastianelli, P.~Benincasa and S.~Giombi,
  ``Worldline approach to vector and antisymmetric tensor fields. II.,''
  JHEP {\bf 0510} (2005) 114
  [hep-th/0510010].
  %%CITATION = HEP-TH/0510010;%%

%\cite{Bastianelli:2004zp}
\bibitem{Bastianelli:2004zp}
  F.~Bastianelli and C.~Schubert,
  ``One loop photon-graviton mixing in an electromagnetic field: Part 1,''
  JHEP {\bf 0502} (2005) 069
  [gr-qc/0412095].
  %%CITATION = GR-QC/0412095;%%

%\cite{Gershun:1979fb}
\bibitem{Gershun:1979fb}
  V.~D.~Gershun and V.~I.~Tkach,
  ``Classical and quantum dynamics of particles with arbitrary spin,''
  JETP Lett.\  {\bf 29} (1979) 288
   [Pisma Zh.\ Eksp.\ Teor.\ Fiz.\  {\bf 29} (1979) 320].
  %%CITATION = JTPLA,29,288;%%

%\cite{Howe:1989vn}
\bibitem{Howe:1989vn}
  P.~S.~Howe, S.~Penati, M.~Pernici and P.~K.~Townsend,
  ``A particle mechanics description of antisymmetric tensor fields,''
  Class.\ Quant.\ Grav.\  {\bf 6} (1989) 1125.
  %%CITATION = CQGRD,6,1125;%%

%\cite{Kuzenko:1995mg}
\bibitem{Kuzenko:1995mg}
  S.~M.~Kuzenko and Z.~.VYarevskaya,
  ``Conformal invariance, N extended supersymmetry and massless spinning particles in anti-de Sitter space,''
  Mod.\ Phys.\ Lett.\ A {\bf 11} (1996) 1653
  [hep-th/9512115].
  %%CITATION = HEP-TH/9512115;%%

%\cite{Bastianelli:2008nm}
\bibitem{Bastianelli:2008nm}
  F.~Bastianelli, O.~Corradini and E.~Latini,
  ``Spinning particles and higher spin fields on (A)dS backgrounds,''
  JHEP {\bf 0811} (2008) 054
  [arXiv:0810.0188 [hep-th]].
  %%CITATION = ARXIV:0810.0188;%%

  %\cite{Dai:2008bh}
\bibitem{Dai:2008bh}
  P.~Dai, Y.~-t.~Huang and W.~Siegel,
  ``Worldgraph approach to Yang-Mills amplitudes from N=2 spinning particle,''
  JHEP {\bf 0810} (2008) 027
  [arXiv:0807.0391 [hep-th]].
  %%CITATION = ARXIV:0807.0391;%%

%\cite{'tHooft:1974bx}
\bibitem{'tHooft:1974bx}
  G.~'t Hooft and M.~J.~G.~Veltman,
  ``One loop divergencies in the theory of gravitation,''
  Annales Poincare Phys.\ Theor.\ A {\bf 20} (1974) 69.
  %%CITATION = AHPAA,A20,69;%%

  %\cite{Christensen:1979iy}
\bibitem{Christensen:1979iy}
  S.~M.~Christensen and M.~J.~Duff,
  ``Quantizing gravity with a cosmological constant,''
  Nucl.\ Phys.\ B {\bf 170} (1980) 480.
  %%CITATION = NUPHA,B170,480;%%

%\cite{deBerredoPeixoto:2001qx}
\bibitem{deBerredoPeixoto:2001qx}
  G.~de Berredo-Peixoto, A.~Penna-Firme and I.~L.~Shapiro,
  ``One loop divergences of quantum gravity using conformal parametrization,''
  Mod.\ Phys.\ Lett.\ A {\bf 15} (2000) 2335
  [gr-qc/0103043].
  %%CITATION = GR-QC/0103043;%%

  %\cite{Bastianelli:2012bn}
\bibitem{Bastianelli:2012bn}
  F.~Bastianelli, R.~Bonezzi, O.~Corradini and E.~Latini,
  ``Effective action for higher spin fields on (A)dS backgrounds,''
  JHEP {\bf 1212} (2012) 113
  [arXiv:1210.4649 [hep-th]].
  %%CITATION = ARXIV:1210.4649;%%

%\cite{Duff:1980qv}
\bibitem{Duff:1980qv}
  M.~J.~Duff and P.~van Nieuwenhuizen,
  ``Quantum inequivalence of different field representations,''
  Phys.\ Lett.\ B {\bf 94} (1980) 179.
  %%CITATION = PHLTA,B94,179;%%


%\cite{Gibbons:1978ac}
\bibitem{Gibbons:1978ac}
  G.~W.~Gibbons, S.~W.~Hawking, M.~J.~Perry,
  ``Path integrals and the indefiniteness of the gravitational action,''  Nucl.\ Phys.\ B {\bf 138} (1978) 141.  %%CITATION = NUPHA,B138,141;%%  %377 citations counted in INSPIRE as of 03 Apr 2013

%\cite{Mazur:1989by}
\bibitem{Mazur:1989by}
  P.~O.~Mazur, E.~Mottola,
  ``The gravitational measure, solution of the conformal factor problem and stability of the ground state of quantum gravity,''  Nucl.\ Phys.\ B {\bf 341} (1990) 187.  %%CITATION = NUPHA,B341,187;%%  %93 citations counted in INSPIRE as of 03 Apr 2013

  %\cite{Kleinert:1999aq}
\bibitem{Kleinert:1999aq}
  H.~Kleinert, A.~Chervyakov,
  ``Reparametrization invariance of path integrals,''  Phys.\ Lett.\ B {\bf 464} (1999) 257  [hep-th/9906156].  %%CITATION = HEP-TH/9906156;%%  %26 citations counted in INSPIRE as of 03 Apr 2013

%\cite{Bastianelli:2000pt}
\bibitem{Bastianelli:2000pt}
  F.~Bastianelli, O.~Corradini, P.~van Nieuwenhuizen,
  ``Dimensional regularization of the path integral in curved space on an infinite time interval,''  Phys.\ Lett.\ B {\bf 490} (2000) 154  [hep-th/0007105].  %%CITATION = HEP-TH/0007105;%%  %20 citations counted in INSPIRE as of 03 Apr 2013

%\cite{Bastianelli:2000nm}
\bibitem{Bastianelli:2000nm}
  F.~Bastianelli, O.~Corradini, P.~van Nieuwenhuizen,
  ``Dimensional regularization of nonlinear sigma models on a finite time interval,''  Phys.\ Lett.\ B {\bf 494} (2000) 161  [hep-th/0008045].  %%CITATION = HEP-TH/0008045;%%  %19 citations counted in INSPIRE as of 03 Apr 2013

%\cite{Bastianelli:2011cc}
\bibitem{Bastianelli:2011cc}
  F.~Bastianelli, R.~Bonezzi, O.~Corradini, E.~Latini,
  ``Extended SUSY quantum mechanics: transition amplitudes and path integrals,''  JHEP {\bf 1106} (2011) 023  [arXiv:1103.3993 [hep-th]].  %%CITATION = ARXIV:1103.3993;%%  %4 citations counted in INSPIRE as of 03 Apr 2013

  %\cite{Bastianelli:1991be}
\bibitem{Bastianelli:1991be}
  F.~Bastianelli,
  ``The path integral for a particle in curved spaces and Weyl anomalies,''
  Nucl.\ Phys.\ B {\bf 376} (1992) 113
  [hep-th/9112035].
  %%CITATION = HEP-TH/9112035;%%

%\cite{Bastianelli:1992ct}
\bibitem{Bastianelli:1992ct}
  F.~Bastianelli and P.~van Nieuwenhuizen,
  ``Trace anomalies from quantum mechanics,''
  Nucl.\ Phys.\ B {\bf 389} (1993) 53
  [hep-th/9208059].
  %%CITATION = HEP-TH/9208059;%%

%\cite{Muller:1997zk}
\bibitem{Muller:1997zk}
  U.~Muller, C.~Schubert, A.~M.~E.~van de Ven,
  ``A closed formula for the Riemann normal coordinate expansion,''  Gen.\ Rel.\ Grav.\  {\bf 31} (1999) 1759  [gr-qc/9712092].  %%CITATION = GR-QC/9712092;%%  %19 citations counted in INSPIRE as of 03 Apr 2013

%\cite{Bastianelli:2000dw}
\bibitem{Bastianelli:2000dw}
  F.~Bastianelli, O.~Corradini,
  ``6-D trace anomalies from quantum mechanical path integrals,''  Phys.\ Rev.\ D {\bf 63} (2001) 065005  [hep-th/0010118].  %%CITATION = HEP-TH/0010118;%%  %14 citations counted in INSPIRE as of 03 Apr 2013

%\cite{Bastianelli:2006rx}
\bibitem{Bastianelli:2006rx}
  F.~Bastianelli, P.~van Nieuwenhuizen,
  ``Path integrals and anomalies in curved space,''
  Cambridge University Press, Cambridge UK (2006).

\end{thebibliography}
\end{document}